\documentstyle[12pt,citesort]{dubna98}
\begin{document}
\newcommand{\eq}{\begin{eqnarray}}
\newcommand{\en}{\end{eqnarray}}
\newcommand{\Ms}{M^\star}
\newcommand{\Ps}{P^\star}
\newcommand{\Es}{E^\star}
\newcommand{\mc}{m_\pi^2}
\newcommand{\mo}{m_{\pi^0}^2}
\newcommand{\dpi}{\Delta_\pi}
\begin{center}
{\bf $\pi^+\pi^-$ ATOM IN CHIRAL THEORIES\\}
\vspace*{0.8cm}
M.A. IVANOV$^1$, V.E. LYUBOVITSKIJ$^{1,2}$, E.Z. LIPARTIA$^{3,4}$ AND A.G. RUSETSKY$^{1,4}$\\
\vspace*{0.2cm}
{\it $^1$ Bogoliubov Laboratory of Theoretical Physics,\\}
{\it Joint Institute for Nuclear Research, 141980, Dubna, Russia\\}
{\it $^2$ Department of Physics,}
{\it Tomsk state University, 634050 Tomsk, Russia\\}
{\it $^3$ Laboratory for Computational Technique and Automation,\\}
{\it Joint Institute for Nuclear Research, 141980, Dubna, Russia\\}
{\it $^4$ Institute of High Energy Physics,}
{\it Tbilisi State University, 380086 Tbilisi, Georgia}
\end{center}

\vspace*{.5cm}
\begin{abstract}
{\small
We review the existing theoretical approaches to the study of the
observables of hadronic atoms. In the relativistic perturbative
approach based on the Bethe-Salpeter equation and Chiral Perturbation Theory
we derive the general expression for the $\pi^+\pi^-$ atom lifetime.
The lowest-order correction to the relativistic
Deser-type formula for the atom width is predicted to be
$(6.1\pm 3.1)\%$.
}
\end{abstract}

\vspace*{1cm}
The study of the pion-pion scattering which forms one of the basic
building blocks in the hierarchy of strong interaction processes, enables one
to gain a deeper insight in the nature of strong interactions.
According to common belief, low-energy interactions
of pions are described within the Chiral Perturbation Theory (ChPT)~\cite{ChPT}
which exploits a full content of global QCD symmetries.
In the standard scheme with a "large" condensate, the $S$-wave
$\pi\pi$ scattering lengths are predicted
to be $a_0^0=0.217$ and $a_0^0-a_0^2=0.258$ in units of the inverse
charged pion mass~\cite{Bijnens}. The calculations within the Generalized
ChPT with a small quark condensate which contain more parameters,
lead to a most likely value $a_0^0=0.27$~\cite{Stern}.
Despite a significant difference between these numbers, both
results for the scattering length $a_0^0$ are compatible with the
experimental value $a_0^0=0.26\pm 0.05$~\cite{Pocanic}. Consequently, a
precise measurement of $\pi\pi$ scattering lengths will be an excellent
test of the ChPT.

The existing experimental information on the $S$-wave $\pi\pi$
scattering lengths is extracted either from the data on the process
$\pi N\rightarrow\pi\pi N$ or from the study of $K_{e4}$
decay (for the review of the recent status of $\pi\pi$ experiments see,
e.g.~\cite{Pocanic}). The forthcoming DIRAC experiment PS212 at CERN
is aimed at the high-precision measurement of the lifetime of the
$\pi^+\pi^-$ atom~\cite{DIRAC}. This will allow for the determination of the
quantity $a_0^0-a_0^2$ with an accuracy of $5\%$ and thus will provide
a decisive probe of the predictions of ChPT. The experiments on the
measurement of the hadronic atom observables are carried out also
at PSI, KEK, Frascati, Uppsala.

The experimental study of $\pi^+\pi^-$ atom characteristics provides a direct
information about the strong $\pi\pi$ amplitudes near threshold since the
average momenta ($\sim \alpha m_\pi$) of constituents in the atom
is much less than either of two typical scales characterizing strong
interactions in chiral theories
($\sim m_\pi$ and $4\pi F_\pi\sim 1~GeV$).
The huge difference between the bound-state and
strong scales leads to a clear-cut factorization of strong and
electromagnetic interactions in the observables of the pionium which is
already explicit in the lowest-order Deser formula for the
ground-state atom lifetime~\cite{Deser}
\eq\label{DESER}
\tau_1^{-1}=\frac{16\pi}{9}\,
\biggl(\frac{2\,\Delta m_\pi}{m_\pi}\biggr)^{{1}/{2}}
(a_0^0-a_0^2)^2\,\,|\Psi_1(0)|^2
\en
where $\Psi_1(0)$ denotes the value of the nonrelativistic Coulomb
ground-state wave function at the origin. Note that the Coulomb interaction
which is mainly responsible for the atom formation, and strong interactions
which lead to its decay, enter separately into the expression of the
lifetime through the Coulomb wave function and the scattering lengths,
respectively.
However, in the view of the
forthcoming DIRAC high-precision experiment it is important to evaluate
corrections to this formula in order to provide a careful determination
of the strong scattering lengths. In particular, one can readily observe
from Eq.~(\ref{DESER}) that the change of the "reference mass" from the
charged to the neutral pion mass in the definition of strong scattering
lengths leads to a $\sim 12\%$ variation in the lifetime. In order to avoid
this ambiguity one has to perform a choice of the "reference mass" in
the ideal, isotopically symmetric world and evaluate all mass shift corrections
which emerge due to the $m_{\pi^\pm}-m_{\pi^0}$ mass difference in the
pionium observables. All other corrections of an electromagnetic or a
strong origin should also be calculated.

Historically, the first
attempts to calculate the corrections to the pionium lifetime
within the field-theoretical approaches were
focused on the evaluation of the quantities entering into Eq.~(\ref{DESER}):
the scattering lengths and the wave function at the origin. The value of
$a_0^0-a_0^2$ was calculated in various low-energy phenomenological
models~\cite{Efimov,Belkov,Volkov}; The $\pi\pi$ strong interaction potential
was constructed and the corrections to the pionium energy
and the value of the wave function at the origin due to
strong interactions, vacuum polarization, finite size of pion were
evaluated, taking into account only the discrete Coulomb
spectrum~\cite{Efimov}; The vacuum polarization effect in the pionium
observables was studied in the quasipotential approach~\cite{Faustov};
Retardation correction to the pionium lifetime was calculated on the
basis of Bethe-Salpeter (BS) approach~\cite{Silagadze}; Electromagnetic
radiative corrections to the $\pi\pi$ scattering lengths and the pionium
lifetime were evaluated~\cite{Kuraev}. Note however that the consistent
approach to the calculation of bound-state characteristics can not
be confined solely to the quantities entering into the Deser formula.
Instead, to this end a systematic field-theoretical framework is needed which
is free of any double counting problems. In the several recent papers
an attempt was made to carry out a comprehensive field theoretical analysis
of the pionium problem. Namely, in~\cite{Sazdjian1,Sazdjian2} on the basis of the
3-dimensional constraint theory bound-state equations the mass shift,
radiative and second-order strong corrections have been evaluated.
In our previous papers~\cite{Atom1,Atom2,Atom3} a systematic perturbative approach based
on the BS equation is constructed, and a complete set of the
lowest-order corrections to the pionium lifetime is evaluated. The vacuum
polarization effect on the pionium observables is evaluated in~\cite{Labelle}
within the nonrelativistic QED.

In a nonrelativistic potential approach to the problem~\cite{Rasche}
the expression for the lifetime, including the corrections, is written
in terms of physical scattering lengths in charged and neutral channels
in the presence of Coulomb interactions. This is a counterpart of the
lowest-order Deser-type factorization: the bound-state observables
are expressed in terms of the quantities characterizing the scattering
process, and these quantities differ from the purely strong ones by
a small and calculable mass shift and electromagnetic corrections.
In order to evaluate these corrections, one needs the $\pi\pi$
potential to be given explicitly. Assuming local and energy-independent
strong potentials in the isospin symmetry limit, one reproduces the
$\pi\pi$ scattering phase shifts calculated from ChPT. At the next step
the Coulomb potential is added and the mass shift effect is
included via assigning the physical masses to pions, the strong potential
remaining unchanged. The sign of the largest mass shift correction
in the scattering
lengths and the pionium lifetime calculated in this manner~\cite{Rasche}
turns out to be opposite to that emerging from the field-theoretical
calculations~\cite{Sazdjian1,Sazdjian2,Atom3}, provided the charged pion mass is
chosen to be the "reference mass". Owing to the derivative character
of pion couplings in the chiral Lagrangian, a possible reason for this
discrepancy might be an explicit energy independence of strong potentials
used in these calculations.

To summarize, one needs an accurate theoretical estimate
of the lifetime of the pionium which will be measured in a high
precision DIRAC experiment. The underlying dynamics of pions should
be described by the Lagrangian of ChPT in order to provide a reliable
and unambiguous test of the predictions of ChPT.

In our previous papers~\cite{Atom1,Atom2,Atom3} we have formulated a consistent
field-theoretical approach to the calculation of the pionium observables.
Below the basic ideas and assumptions of our approach are given.

The pions within our approach
are described by the elementary fields in the Lagrangian.
This allows one to write down the bound-state BS equation for the
metastable $\pi^+\pi^-$ atom. The necessary link to the strong amplitudes
from ChPT where the pions are described by the pseudoscalar quark densities,
proceeds via the Deser-type factorization in analogy with the potential
theory. Namely, we express all corrections to the Deser formula in terms
of the on-mass-shell pion amplitudes which are further identified with
the amplitudes from ChPT.

In our calculations we always adopt the so-called "local" approximation
which consists is suppressing the relative momentum dependence of the
strong amplitudes. The origin of this approximation can be traced back
to the huge difference between the bound-state and strong momentum scales.
Since the tree-level strong amplitude for the process
$\pi^+\pi^-\to \pi^0\pi^0$ does not depend on the relative momentum
of the pions, the corrections arising due to this approximation,
by a simple power counting, should be suppressed by a product of {\it two}
small factors:
the fine structure constant $\alpha$ which is the ratio of the bound-state
momentum and the pion mass, {\it and} the factor
$m_\pi/(4\pi F_\pi)\sim 10^{-1}$ being the ratio of two strong scales in
chiral theories. For this reason we at the present stage completely
neglect the corrections coming from using the "local" approximation.

The pionium in our approach is described by the exact bound-state
BS equation
\eq\label{BS}
<\chi|G_0^{-1}(P)=<\chi|V(P)
\en
where $\chi$ denotes the BS wave function of pionium, $G_0$ is the
free Green's function of the $\pi^+\pi^-$ pair and $V$ is the full BS
kernel which apart from all two-particle irreducible diagrams includes
the self-energy corrections in two outgoing charged pion legs~\cite{Atom1,Atom2,Atom3}.
The square of the c.m. momentum $P$ takes the value
$P^2=\bar M^2=M^2-iM\Gamma$ where $M$ denotes the "mass" of an atom,
and $\Gamma$ stands for the decay width.

In order to perform the perturbative expansion of the bound-state
observables we single out from the full kernel $V$ the instantaneous
Coulomb part $V_C$ which
is responsible for the formation of the bound state composed of
$\pi^+$ and $\pi^-$. The "remainder" of the potential which is treated
perturbatively is denoted by $V'=V-V_C$.

The "unperturbed" part of the full kernel $V_C$ is chosen in the
form guaranteeing the exact solvability of the corresponding BS equation.
The solution of this equation yields the relativistic Coulomb wave function
$\psi_C$ with the corresponding eigenvalue
${\Ps}^2={\Ms}^2=m_\pi^2(4-\alpha^2)$
where $m_\pi$ is the charged pion mass. The solution of the full BS equation
is expressed in terms of the unperturbed solution as follows~\cite{Atom2,Atom3}
\eq\label{wf-connection}
<\chi|={\rm const}\times
<\psi_C|\bigl[1+(G_0^{-1}(P)-G_0^{-1}(\Ps)-V'(P))G_RQ\bigr]^{-1}
\en
where $G_R$ denotes the pole-subtracted part of the exact relativistic
Coulomb Green's function and $Q$ is the projection operator on the
subspace orthogonal to the unperturbed ground-state BS wave function~\cite{Atom2,Atom3}.
Substituting this solution into the complete BS equation (\ref{BS})
we arrive at the final relation
$$
<\psi_C|\bigl[ 1+(G_0^{-1}(P)-G_0^{-1}(\Ps)-V'(P)) G_R Q\bigr]^{-1}
(G_0^{-1}(P)-G_0^{-1}(\Ps)-V'(P))|\psi_C>=0
$$
which serves as the basic equation for performing the perturbative expansion
of the bound-state observables. The explicit form of the operator $G_RQ$
is given and the perturbation kernel $V'$ is known up to any given
order of loop expansion. The only unknown quantities entering (parametrically)
into this expression are the real and imaginary parts of $P^2$.
Consequently, expanding this expression in powers of $V'$, we arrive at the
recursive relations which define the mass and width of the bound state
in any given order.

In order to provide the necessary Deser-type
factorization in the obtained results, one needs a further classification
of diagrams entering into the definition of $V'$. We consider the
following subsets of diagrams:

\noindent
1. A purely strong part, which is isotopically invariant. This part
survives when electromagnetic interactions are "turned off" the theory.

\noindent
2. The part which is responsible for the $m_{\pi^\pm}-m_{\pi^0}$
electromagnetic mass difference

\noindent
3. Remaining electromagnetic effects, including the exchanges of
virtual photons.

The part 1 includes strong interactions which govern the
decay of a pionium. The part 2 makes this decay kinematically allowed.
Consequently, it is natural to consider them together,
denoting the corresponding potential as $V_{12}=V_1+V_2$. The $T$-matrix
corresponding to summation of the potential $V_{12}$ in all orders
is given by $T_{12}(P)=V_{12}(P)+V_{12}(P)G_0(P)T_{12}(P)$. The rest
of the potential is referred to as $V_3=V'-V_{12}$ and is treated
perturbatively. Further, the quantity $G_RQ$ in our basic relation can
be written as $G_RQ=G_0(\Ps)+\delta G$ where $\delta G$ corresponds to
the exchange of the ladder of Coulomb photons and is also considered
as a perturbation. Then in the first order in $V_3$ and $\delta G$ the basic
relation takes the form
\eq\label{LARGE}
&&0=-2i\Ms\delta M-<\psi_C|T_{12}|\psi_C>+\nonumber\\[2mm]
&&+\,\delta M<\psi_C|(G_0^{-1})'G_0T_{12}|\psi_C>
+(\delta M)^2<\psi_C|(1+T_{12}G_0)(G_0^{-1})''|\psi_C>+
\nonumber\\[2mm]
&&+<\psi_C|(\delta M(G_0^{-1})'-T_{12})\delta GT_{12}|\psi_C>
-<\psi_C|(1+T_{12}G_0)V_3(1+G_0T_{12})|\psi_C>\quad\quad
\en
where $G_0=G_0(\Ms)$ and the prime stands for the differentiation with
respect to $\Ms$.

In the "local" approximation $T_{12}$ does not depend on the relative
momenta and thus can be taken out of the matrix elements.
Neglecting first the
corrections (only the first line of Eq.~(\ref{LARGE}) contributes),
we arrive at the relativistic counterpart of Eq.~(\ref{DESER})
\eq\label{rel-deser}
\Delta E^{(1)}={\mbox{Re}}\biggl( \frac{iT_{12}}{2\Ms m_\pi}\phi_0^2\biggr),
\quad\quad
-\frac{1}{2}\Gamma^{(1)}={\mbox{Im}}\biggl(\frac{iT_{12}}{2\Ms m_\pi}\phi_0^2\biggr)
\en
\eq
{\mbox{Re}}\,(iT_{12})&=&
16\pi\,{\cal T}_{\pi^+\pi^-\rightarrow\pi^+\pi^-}(4m_\pi^2;\vec 0,\vec 0\,)
\label{REAL}
\\[2mm]
{\mbox{Im}}\,(iT_{12})&=&
-16\pi\,\biggl(\frac{\Delta m_\pi}{2m_\pi}\biggr)^{1/2}
\biggl( 1-\frac{\Delta m_\pi}{2m_\pi}\biggr)^{1/2}
\,\,|{\cal T}_{\pi^+\pi^-\rightarrow\pi^0\pi^0}(4m_\pi^2;\vec 0,\vec q_0\,)|^2,\quad
\label{IMAGINARY}
\en
and ${\cal T}(s;\vec p,\vec q\,)$ denote the (dimensionless) $S$-wave
$\pi\pi$ strong scattering amplitudes which include the effect of
$m_{\pi^\pm}-m_{\pi^0}$ mass difference. $\vec q_0$ is the relative momentum
of the $\pi^0\pi^0$ pair at the threshold $s=4m_\pi^2$, with the magnitude
given by the relation $m_\pi^2=m_{\pi^0}^2+\vec q_0^{~2}$.

In order to calculate the corrections to the relativistic Deser formula,
one has to evaluate the integrals entering into the Eq.~(\ref{LARGE}).

\noindent {\bf 1.}
The correction due to the shift of the bound-state pole by strong
interactions is determined by two terms in the second line of Eq.~(\ref{LARGE}).
Calculating these integrals explicitly and using Eqs.~(\ref{rel-deser}) we
obtain
\eq
\delta_S=-\frac{9}{8}\,\frac{\Delta E^{(1)}}{E_1}=
-5.47\times 10^{-3}\,\,m_\pi(2a_0^0+a_0^2)
\en
where $E_1$ denotes the unperturbed ground-state binding energy.

\noindent {\bf 2.}
The correction due to the relativistic modification of the Coulomb wave
function is given by the value of the wave function $\psi_C$ at the origin
which for the particular choice of the kernel used in the calculations
is related to its nonrelativistic counterpart by
\eq
\psi_C(0)=\Psi_1(0)\,\, (1-0.381\alpha+\cdots)
\en
and the correction in the decay width which is proportional
to $|\psi_C(0)|^2$
is twice as large.

\noindent {\bf 3.}
The correction due to the exchange of Coulomb photon ladders is given
by the first term in the third line of Eq.~(\ref{LARGE}) containing
$\delta G$. Using the known expression for the relativistic Coulomb
Green's function~\cite{Atom2,Atom3} and evaluating this integral explicitly,
we obtain
\eq
\delta_C=(1/2+2.694-{\rm ln}\alpha)\,\Delta E^{(1)}/E_1=
3.95\times 10^{-2}\,\,m_\pi(2a_0^0+a_0^2)
\en

\noindent {\bf 4.}
In order to obtain the mass shift and radiative corrections, in
the perturbation kernel $V_3$ one has to take into account
the residual photon exchange diagram and the self-energy corrections
in the outgoing charged pion legs. The net result of this effect consists
in the cancellation of the correction due to the relativistic
modification of the Coulomb wave function
and in replacing the quantity ${\cal T}_{\pi^+\pi^-\rightarrow\pi^0\pi^0}$
in Eq.~(\ref{IMAGINARY}) by the Coulomb pole removed
full amplitude for the process $\pi^+\pi^-\rightarrow\pi^0\pi^0$.
Using the expression of this amplitude~\cite{Knecht}, and isolating the purely
strong isotopically symmetric amplitude with a common mass equal to
the charged pion mass, we arrive at the following expressions for the
mass shift and radiative corrections
\eq
\delta_M=\frac{2\Delta m_\pi^2}{3m_\pi^2}
\biggl(1+\frac{m_\pi^2}{96\pi^2F_\pi^2}
\biggl(48+\frac{16}{3}\bar l_1-\frac{16}{3}\bar l_2+7\bar l_3-36\bar l_4
\biggr)\biggr)
\en
\eq
\delta_{em}=\frac{\alpha}{12\pi}
(-30+3{\cal K}_1^{\pm 0}-{\cal K}_2^{\pm 0})
\en
where ${\cal K}_i^{\pm 0}$ and $\bar l_i$ denote the low-energy constants
of ChPT~\cite{ChPT,Knecht}.

\noindent {\bf 5.}
The correction due to the vacuum polarization is of order
$O(\alpha^2m_\pi/m_e)$ (the contribution due to the hadronic vacuum
polarization is negligibly small). Taking into account the
corresponding diagram in the kernel $V_3$ and evaluating numerically
the resulting integral, we arrive at the following result
\eq
\delta_{vac}=\frac{3}{16}\,\alpha^2\,\frac{m_\pi}{m_e}\times 0.6865
\en
which completely agrees with the zero-Coulomb piece of the result
given in Ref. \cite{Labelle}. The additional contributions coming
from one-Coulomb and many-Coulomb pieces which are also considered
in this paper, within our approach arise in the second-order of the
perturbation theory.

\noindent {\bf 6.}
The correction to the pionium decay width due to the finite size
effect is caused by the modification of the instantaneous Coulomb
interaction by the pion loop in the two pion-photon vertex (the
corresponding diagram is contained in the kernel $V_3$). For simplicity,
we approximate this vertex by a monopole parameterization.
After calculating the corresponding matrix element explicitly, we obtain
\eq
\delta_F=\frac{2\alpha}{3\pi}\,\,\mc<r^2>_V^\pi\,\,
{\rm ln}\,\biggl(\frac{1}{24}\,\mc<r^2>_V^\pi\biggr)
\en
where $<r^2>_V^\pi$ denotes the square charge radius of the pion.

The relativistic Deser formula with all above corrections takes the form
$$
\tau_1^{-1}=\frac{16\pi}{9}\,
\biggl(\frac{2\,\Delta m_\pi}{m_\pi}\biggr)^{{1}/{2}}
\biggl(1-\frac{\Delta m_\pi}{2m_\pi}\biggr)^{1/2}
(a_0^0-a_0^2)^2\,\,\phi_0^2\,\,
(1+\delta_S+\delta_C+\delta_M+\delta_{em}+\delta_{vac}+\delta_F)
$$
where $a_0^0$ and $a_0^2$ denote the $\pi\pi$ scattering lengths
in the isospin-symmetric case, with the charged pion mass taken
to be the common mass of the pion isotriplet.

In our calculations for the constants $\bar l_i$ we take the
numerical values from Ref. \cite{ChPT}
$\bar l_1=-2.3\pm 2.7$, $\bar l_2=6.0\pm 1.3$, $\bar l_3=2.9\pm 2.4$,
$\bar l_4=4.3\pm 0.9$. Also, we use the values
$\frac{e^2F_\pi^2}{\mc}\,{\cal K}_1^{\pm 0}=1.8\pm 0.9$,
$\frac{e^2F_\pi^2}{\mc}\,{\cal K}_2^{\pm 0}=0.5\pm 2.2$
from Ref.~\cite{Knecht}.
Other input parameters in our calculations are the S-wave $\pi\pi$
scattering lengths: $a_0^0=0.217m_\pi^{-1}$, $a_0^2=-0.041m_\pi^{-1}$
calculated in ChPT and the e.m. charge radius of pion
$<r^2>_V^\pi=0.439~Fm$ \cite{ChPT}. The results of our calculations
are shown in Table 1.

As we observe from Table 1, the largest correction in the decay width
is caused by the mass shift effect.
Our result for the sum of mass shift and electromagnetic radiative
corrections generally agrees in sign and magnitude
with the recent field-theoretical calculations from Ref. \cite{Sazdjian2}.
The sign of the mass shift effect obtained in the nonrelativistic
scattering theory approach \cite{Rasche} turns out to be opposite
as compared to our result, and is of the same order of magnitude.

Our result for the correction due to the exchange of Coulomb photons
agrees with the result obtained in the potential scattering theory
\cite{Rasche}
and disagrees with the result from Refs. \cite{Sazdjian1,Sazdjian2}.
Numerically the largest part in this
effect comes from the nonanalytic ${\rm ln}\alpha$ piece which is
exactly the same in our approach and in the potential theory, and
is absent in Refs. \cite{Sazdjian1,Sazdjian2}.

Thus, we have evaluated a complete set of the lowest-order corrections to
the pionium decay width. For a full understanding of the problem,
however, the reason for the difference in sign in the mass shift effect
in the field-theoretical and potential theories should be investigated in
detail. Also, one should have a reliable quantitative estimate of the
accuracy of "local" approximation including a proper treatment of arising
new UV divergences.

\newpage
\noindent {\bf Table 1.} Corrections to the $\pi^+\pi^-$ atom decay width.
\begin{center}
\def\arraystretch{1.0}
\begin{tabular}{|l|c|c|}
\hline
~Effect                    & ~Value~        & ~Correction (in \%)~  \\
\hline
~Strong                    & $\delta_S$     & $-0.22$               \\
\hline
~Coulomb photon exchange~  & $\delta_C$     & $+1.55$               \\
\hline
~Mass shift                & $\delta_M$     & $+2.99\pm 0.77$       \\
\hline
~Electromagnetic radiative & $\delta_{em}$  & $+1.73\pm 2.31$       \\
\hline
~Vacuum polarization       & $\delta_{vac}$ & $+0.19$               \\
\hline
~Finite size               & $\delta_F$     & $-0.16$               \\
\hline
~Total                     & $\delta_{tot}$ & $+6.1\pm 3.1$         \\
\hline
\end{tabular}
\end{center}

\vspace*{0.2cm}

{\it Acknowledgments}.
We thank V. Antonelli, A. Gall, A. Gashi, J. Gasser,
E.A. Kuraev, H. Leutwyler, P. Minkowski, L.L. Nemenov, E. Pallante, H. Sazdjian
and Z. Silagadze
for useful
discussions, comments and remarks. A.G.R. thanks Bern University for
the hospitality where part of this work was completed.
This work was supported in part by the Russian Foundation for
Basic Research (RFBR) under contract 96-02-17435-a.

\end{document}